\documentclass[a4paper,fleqn]{cas-dc}

\usepackage[numbers,authoryear,longnamesfirst]{natbib}
\hypersetup{linkcolor=blue}
\usepackage{lineno}
\usepackage{ulem}
\def\tsc#1{\csdef{#1}{\textsc{\lowercase{#1}}\xspace}}
\tsc{WGM}
\tsc{QE}

\begin{document}

\let\WriteBookmarks\relax
\def\floatpagepagefraction{1}
\def\textpagefraction{.001}

\shorttitle{The stability of unevenly spaced planetary system}    

\shortauthors{Yang et al.}  

\title [mode = title]{The stability of unevenly spaced planetary systems}  

\author[1]{Sheng Yang}[orcid=0009-0004-2783-7377]

\fnmark[1]

\cormark[1]

\ead{shengyang@mail.sdu.edu.cn}

\affiliation[1]{organization={School of Physics, Shandong University},
            addressline={27 Shandanan Road}, 
            city={Jinan},
            postcode={250100}, 
            state={Shandong},
            country={China}}

\author[2]{Liangyu Wu}[orcid=0009-0003-2402-0735]

\fnmark[1]

\affiliation[2]{organization={School of Physics and Astronomy, Shanghai Jiao Tong University},
            addressline={800 Dongchuan Road}, 
            city={Shanghai},
            postcode={200240}, 
            country={China}}

\author[3]{Zekai Zheng}[orcid=0000-0003-0483-5251]

\affiliation[3]{organization={School of Physics, Zhejiang University},
            addressline={866 Yuhangtang Road}, 
            city={Hangzhou},
            postcode={310058}, 
            state={Zhejiang},
            country={China}}

\author[2,4]{Masahiro Ogihara}[orcid=0000-0002-8300-7990]

\affiliation[4]{organization={Tsung-Dao Lee Institute, Shanghai Jiao Tong University},
            addressline={520 Shengrong Road}, 
            city={Shanghai},
            postcode={201210}, 
            country={China}}

\affiliation[5]{organization={School of Geophysics and Information Technology, China University of Geosciences (Beijing)},
            addressline={29 Xueyuan Road}, 
            city={Beijing},
            postcode={100083}, 
            country={China}}
\author[4]{Kangrou Guo}[orcid=0000-0001-6870-3114]
\author[5]{Wenzhan Ouyang}[orcid=0009-0002-6630-3189]
\author[4]{Yaxing He}[orcid=0000-0001-5264-1924]

\cortext[1]{Correspondence to: School of Physics,Shandong University, 27 Shandanan Road, Jinan, 250100, Shandong, China}

\fntext[1]{Both authors have contributed equally.}

\begin{abstract}
Studying the orbital stability of multi-planet systems is essential to understand planet formation, estimate the stable time of an observed planetary system, and advance population synthesis models. 
Although previous studies have primarily focused on ideal systems characterized by uniform orbital separations, in reality a diverse range of orbital separations exists among planets within the same system.
This study focuses on investigating the dynamical stability of systems with non-uniform separation.
We considered a system with 10 planets with masses of $10^{-7}$ solar masses around a central star with a mass of $1$ solar mass.
We performed more than 100,000 runs of \textit{N}-body simulations with different parameters.
Results demonstrate that reducing merely one pair of planetary spacing leads to an order of magnitude shorter orbital crossing times that could be formulated based on the Keplerian periods of the closest separation pair. 
Furthermore, the first collisions are found to be closely associated with the first encounter pair that is likely to be the closest separation pair initially.
We conclude that when estimating the orbital crossing time and colliding pairs in a realistic situation, updating the formula derived for evenly spaced systems would be necessary.

\end{abstract}


\maketitle

\section{Introduction} \label{sec:intro}
Dynamical stability of multiplanetary systems is important from several perspectives. At the end of the runaway and oligarchic growth phases, protoplanets with masses of approximately 0.1--1 Earth masses are formed, maintaining certain orbital separation \citep[e.g.,][]{kokubo_oligarchic_1998}.
Collisions and scattering resulting from the orbital instability of protoplanets affect the orbital configuration of planetary systems. Furthermore, an improved understanding of orbital stability (e.g., orbital crossing time) can be leveraged in planetary population synthesis models. In addition, exoplanet observations suggest that most exoplanet systems have multiple planets \citep[e.g.,][]{lissauer_architecture_2011}.
Studying dynamical instability can also be effective in estimating the stability of such exoplanet systems.

The orbital instability of planetary systems has been studied from different perspectives using \textit{N}-body simulations.  
Previous research has explored factors such as the number, mass, spatial distribution, eccentricity, and inclination of the planets. 
For instance, \citet{chambers_stability_1996} investigated coplanar planetary systems with uniform spacing and established a relationship between the orbital crossing time and the orbital separation as $\log t_{\rm cross} = b\Delta + c$, where $\Delta$ is the initial separation of the planets in the unit of their mutual Hill radii.
\citet{yoshinaga_stability_1999} explored the influence of initial eccentricity and inclination of the planets on the crossing time: an increase in eccentricity and inclination is found to decrease the orbital crossing time. 
Similarly, \citet{zhou_post-oligarchic_2007} studied these systems and provided extended relations regarding coefficients $b$ and $c$, the function of eccentricity and inclination. 
\citet{smith_orbital_2009} examined closely spaced planetary systems and confirmed that the exponential law relationship between spatial distribution and crossing time still holds.
In addition to these numerical studies, analytical studies have been conducted to understand the dynamical stability \citep[e.g.,][]{2011MNRAS.418.1043Q,petit_path_2020}.

Previous studies have mainly focused on systems with uniform spacing. However, variability in the orbital separation between planets within a system is typically observed in the results of \textit{N}-body simulation of planet formation under the influence of planet-disk interaction \citep[e.g.,][]{izidoro_breaking_2017,ogihara_formation_2018}. Such variability can influence the stability timescale. \citet{pu_spacing_2015} conducted simulations based on different mean and dispersion values of the orbital separation between planets and suggested that the closest pair of planets probably determines the crossing time (see also \citet{wu_dynamical_2019}). In this study, we focus on the impact of a pair of adjacent planets with a narrower separation on the orbit crossing time, without altering the separation of other planets. Furthermore, we consider the pair that experiences the first orbital instability, including close encounter and collisions. 
The purpose of this paper is to observe the role of the closest separation pair on dynamical stability, and a general formulation of orbit stability time should be made in another study.

\section{Methods} \label{sec:methods}

We followed basically the same setup as \citet{chambers_stability_1996}.
Ten planets with a mass of $M=10^{-7} M_{\odot}$ were considered to orbit around a star with a mass of $M_{\odot}$. The orbits of all planets were initially circular and coplanar with the semimajor axis set as $a_{i+1}-a_i= f \Delta R_{{\rm H},i}$, where 
$f$ and $\Delta$ are free parameters for the initial orbital separation, $i=1,2,...9$ and $R_{{\rm H},i}$ is the mutual Hill radius of the planet pair defined as
\begin{flalign}
    &&
    R_{{\rm H},i} = \left(\frac{M_i + M_{i+1}}{3M_\odot}\right)^{1/3} \frac{a_i+a_{i+1}}{2}.
    &&
\end{flalign}
The innermost planet was initially located at $a_1=1 {\rm \,au}$. The initial longitude of the planets were set randomly but the longitude difference between the adjacent planets was larger than 20°.
In this study, we simulated the case where only the orbital separation of one planet pair was reduced unless otherwise noted. In other words, the scaling factor $f$ was less than 1 for one pair of adjacent planets, and $f=1$ for the other pairs. 
The closest separation pair was changed in different simulations. We varied the value of $\Delta$ and the pair for which the initial separation was reduced in our simulations.

We used the open-source \textit{N}-body code \texttt{REBOUND} \citep{rein_rebound_2012} to simulate the evolution of this system.
For orbital integration, \texttt{WHFAST} integrator with a timestep of 0.05 year was used \citep{rein_whfast_2015}.
We stopped the simulation when the mutual Hill radius of two planets overlapped and recorded the orbit crossing time $t_{\rm cross}$ representing the stable time of the system.
For each model with a given set of parameters, we repeated the simulation 500 times.
In certain simulations, we followed the dynamical instability until the collision, in which case we used \texttt{IAS15} integrator with adaptive time stepping \citep{rein_ias15_2015}.
In this case, the internal density of the planet is assumed to be 3\,g\,cm$^{-3}$, and the overlap of the physical radius is considered as a collision. We performed 500 runs on one set unless otherwise stated.

\section{Results} \label{sec:results}

\begin{figure*}[htbp]
  \centering
   \includegraphics[width=1.0\linewidth]{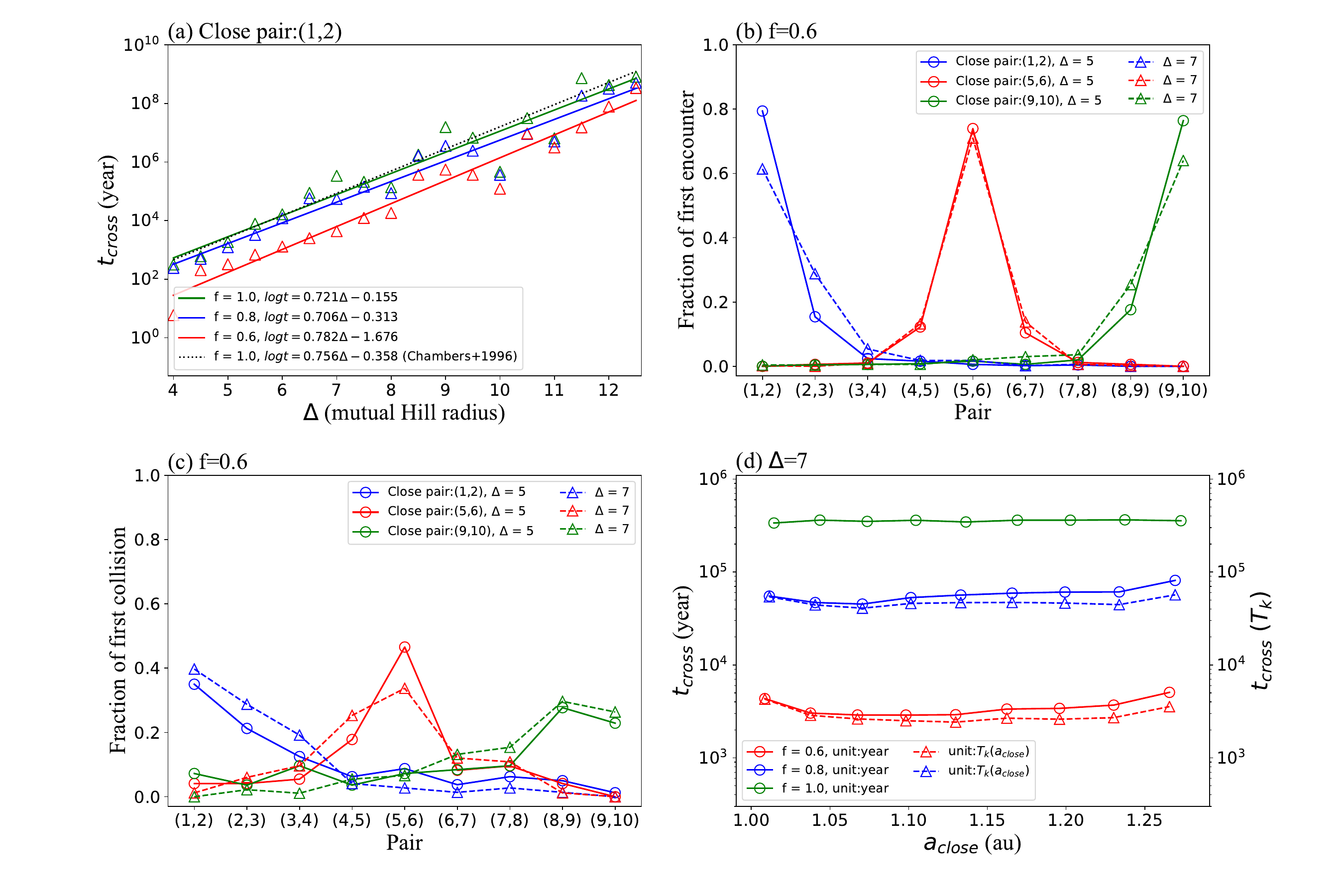}
 \caption{(a) The median orbit crossing time over 500 runs of simulations for different $\Delta$ and $f$. 
 For large separations ($\Delta > 8$), the median of 40 runs is shown.
The closest separation pairs are assumed to be the innermost pairs. The solid lines correspond to least-squares fitting. The fitting of previous study for the uniformly separated case is shown by the black dotted line. (b) Relationship between the first close encounter pair and the initial closest pair for $f=0.6$. (c) Relationship between the first colliding pair and the initial closest pair for $f=0.6$. (d) Relationship between the orbit crossing time of the system and the semimajor axis of the closest pair. The solid line is measured in years, and the dashed line is measured in the Keplerian orbital period of the closest pair.}
    \label{fig:pic1}
\end{figure*}

Figure $\ref{fig:pic1}$(a) demonstrates the median orbit crossing time over 500 runs when the innermost pair has the closest separation. 
Since our approach in this study is to observe the differences from \citet{chambers_stability_1996}, we mainly consider orbital separations similar to those of \citet{chambers_stability_1996} ($\Delta \le 8$). At the same time, the simulation for wider separations ($\Delta > 8$) was also performed as a reference.
The results are confirmed to be consistent with previous studies \citep{chambers_stability_1996} when all planets are initially uniformly separated ($f=1$). In addition to the evident linear relationship between $\log t_{\rm cross}$ and $\Delta$ already known, as the scaling factor $f$ is decreased, systems become dynamically unstable in a shorter period of time. This result implies that when estimating the orbital stability time for unevenly spaced systems, using the orbit crossing time for the evenly spaced case \citep[e.g.,][]{chambers_stability_1996} can overestimate the crossing time by as much as an order of magnitude or more.

Next, we focus on the correlation between the pair that experiences the first close encounter and the initially closest pair, where we define a close encounter event when the separation between two planets causes their mutual Hill radii to overlap. Figure $\ref{fig:pic1}$(b) shows the distribution of probabilities of first close encounters. It is revealed that by decreasing the initial distance between specific adjacent planets, the probability of the first close encounter between these two planets significantly increases and is noticeably higher than the probability of the first close encounter for other planets. This trend does not change with different initial separations (namely $\Delta$=5 or 7).
Furthermore, this correlation is more pronounced for cases with a smaller $f$. 

We examine the relationship between the pairs that first experience close encounters and the pairs that actually collide first. 
In these simulations, we modify the radii of these planets from the Hill radius to the physical radius. Figure $\ref{fig:pic1}$(c) shows the fraction of pairs that experience the first collision. The results show that the pairs that collide due to orbital instability are indeed correlated with the first encounter pair, implying that the pair of planets that has reduced initial separation is more likely to encounter the first collision.

We further studied the change in the orbit crossing time with the change in the closest pair, as shown in Figure $\ref{fig:pic1}$(d). We observe that the orbit crossing time is adequately scaled by year for $f=1$, as in previous studies, for uniformly separated planets at $a\simeq 1\,{\rm au}$ around a solar-mass star \citep{chambers_stability_1996,yoshinaga_stability_1999,zhou_post-oligarchic_2007,smith_orbital_2009}. 
In other words, the orbit crossing time is constant in Figure $\ref{fig:pic1}$(d), although there are slight variations due to the random longitude settings in the initial setup. However, the stability time of the system in years increases as the semimajor axis of the closest pair, $a_{\rm close}$, increases for $f<1$. Here, $a_{\rm close}$ refers to the average of the semimajor axis of two closely spaced planets. This means that care must be taken when expressing the orbit crossing time with the Keplerian orbital period of the innermost planet. Here we find that the orbit crossing time can be again suitably scaled using the Keplerian orbital period with the semimajor axis of $a_{\rm close}$, called $T_{\rm K}(a_{\rm close})$. When $T_{\rm K}(a_{\rm close})$ is used, the crossing time does not increase with increasing $a_{\rm close}$. We also ran similar simulations with different parameters ($f$ and $\Delta$) and arrived at the same conclusion as above\footnote[2]{We further confirmed that this trend was correct by increasing the number of planets to 20 (See Supplementary data).}
\footnote[3]{In addition to the case where only one pair has the closest separation, additional simulations were performed for the case where a second closest pair exists. As a result, we confirmed that the finding that the orbit crossing time is expressed by the orbital periods of the closest separation pair does not change (See Supplementary data).}. In addition, the two ends in Figure $\ref{fig:pic1}$(d) are slightly raised, indicating that the orbit crossing time is slightly longer when the closest separation pair locates the ends. A similar trend is seen in \citet{GRATIA2021114038}, where the eccentricity of one planet is changed for evenly spaced systems.

\section{Summary and discussion} \label{sec:discussion}

We investigate a system consisting of 10 non-uniformly spaced, circular and coplanar planets of equal mass. The system is modified by changing the distance between only one pair of planets. We conducted multiple numerical experiments under each condition with different reduction factors $f$ and initial orbital separation $\Delta$.
The \textit{N}-body simulation results are as follows:
\begin{itemize}
    \item Orbit crossing times become shorter in the case of pairs with shorter initial orbital separations.
    \item There is a correlation between the closest separation pair and the first close encounter or collision.
    \item The orbital crossing time for systems with uneven orbital separation could be expressed based on the orbital periods of the closest separation pair.

\end{itemize}
Note that although we performed many \textit{N}-body simulations with different parameters, the situation we considered is limited; further simulations are needed to see if the result holds for more general systems with continuous variations in orbital separation within a system.

By introducing the orbital crossing time discussed in this study into the planetary population synthesis simulation \citep[e.g.,][]{ida_toward_2013}, the onset of the late dynamical instability phase can be considered without \textit{N}-body simulations. Predicting the stability of discovered exoplanetary systems without time-consuming orbital simulations is also feasible.
The results of this study show that the orbit crossing time can be overestimated by an order of magnitude if the average orbital separation is used to estimate the orbit crossing time. Estimating the orbit crossing time using the orbital separation and the orbital period of the closest pair would be recommended when introducing it into population synthesis calculations and orbital stability calculations for exoplanets.

The results of this study show that the closest separation pair is correlated with the first orbital crossing and collision.
When late dynamical instability occurs near $a=1\,{\rm au}$, orbital crossings occur on a relatively large scale, and all planets in the system are expected to undergo instability and collision. In this case, the first pair to collide may not be extremely important.
However, owing to the large physical radius to Hill radius ratio in close-in orbits ($a<0.1\,{\rm au}$), only local orbital crossings may occur without large-scale orbital crossings throughout the system \citep[e.g.,][]{ogihara_n-body_2009,matsumoto_ejection_2020}.
When considering such situations in population synthesis simulations, local orbital instabilities and collisions must be treated, and the identification of the first colliding pair may be important.

As mentioned above, the general formulation of orbital crossing times is useful for future population synthesis simulations and for estimating the stability of observed exoplanetary systems. Further \textit{N}-body simulations under different conditions should be conducted in the future to develop a general formulation, such as the mass of planets, orbital  inclination, eccentricity, and other such parameters.

\section*{Acknowledgements}
We thank the anonymous referee for constructive comments. This study was initiated at the TDLI Astro Winter Camp 2023.

\appendix
\section*{Appendix A. Supplementary data}
\begin{figure*}[p]
  \centering
   \includegraphics[width=1.0\linewidth]{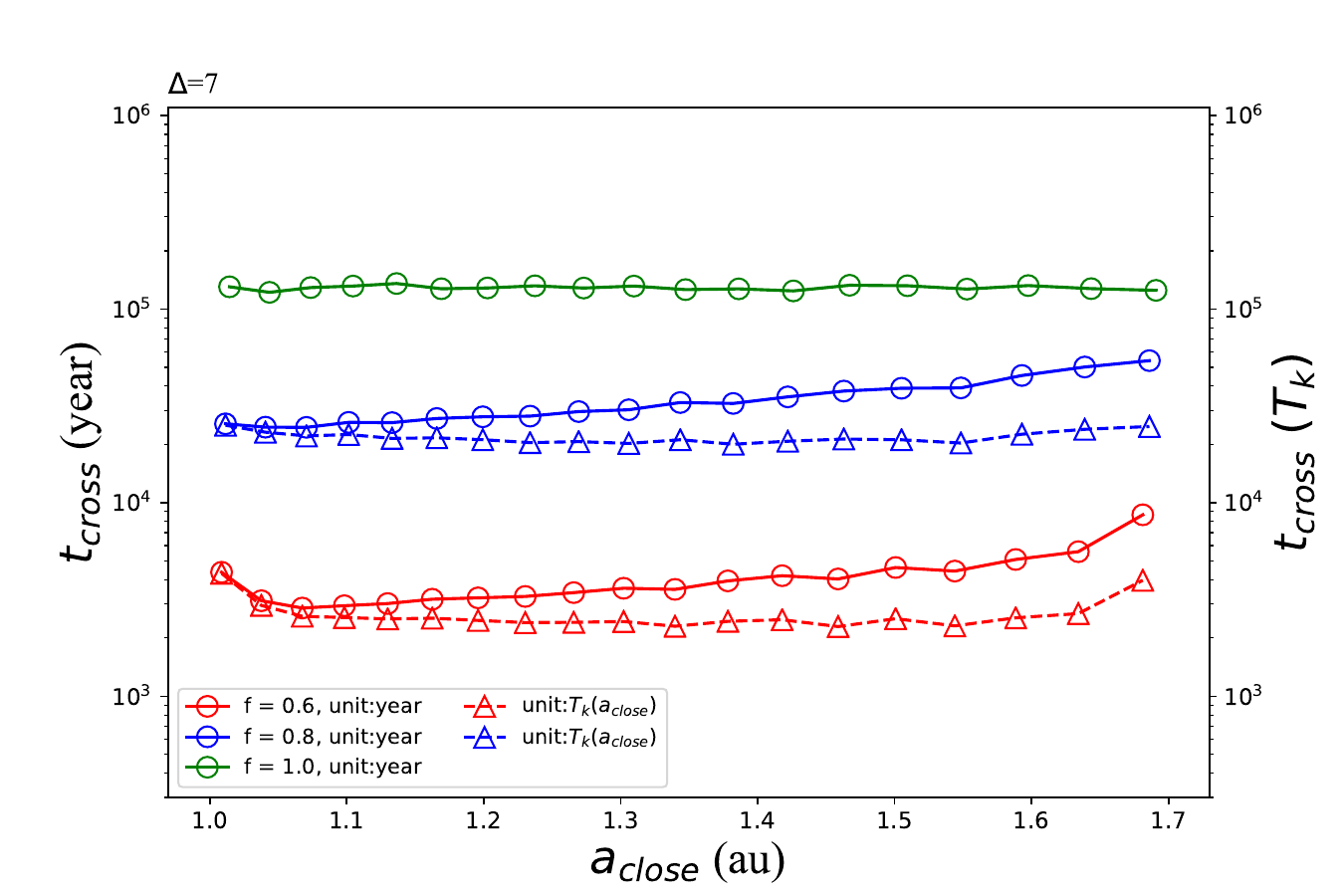}
 \caption{Same as Figure\,1(d) in the main text, but for systems with 20 planets.}
\label{fig:pic_supp_1}
\end{figure*}

\begin{figure*}[H]
  \centering
   \includegraphics[width=1.0\linewidth]{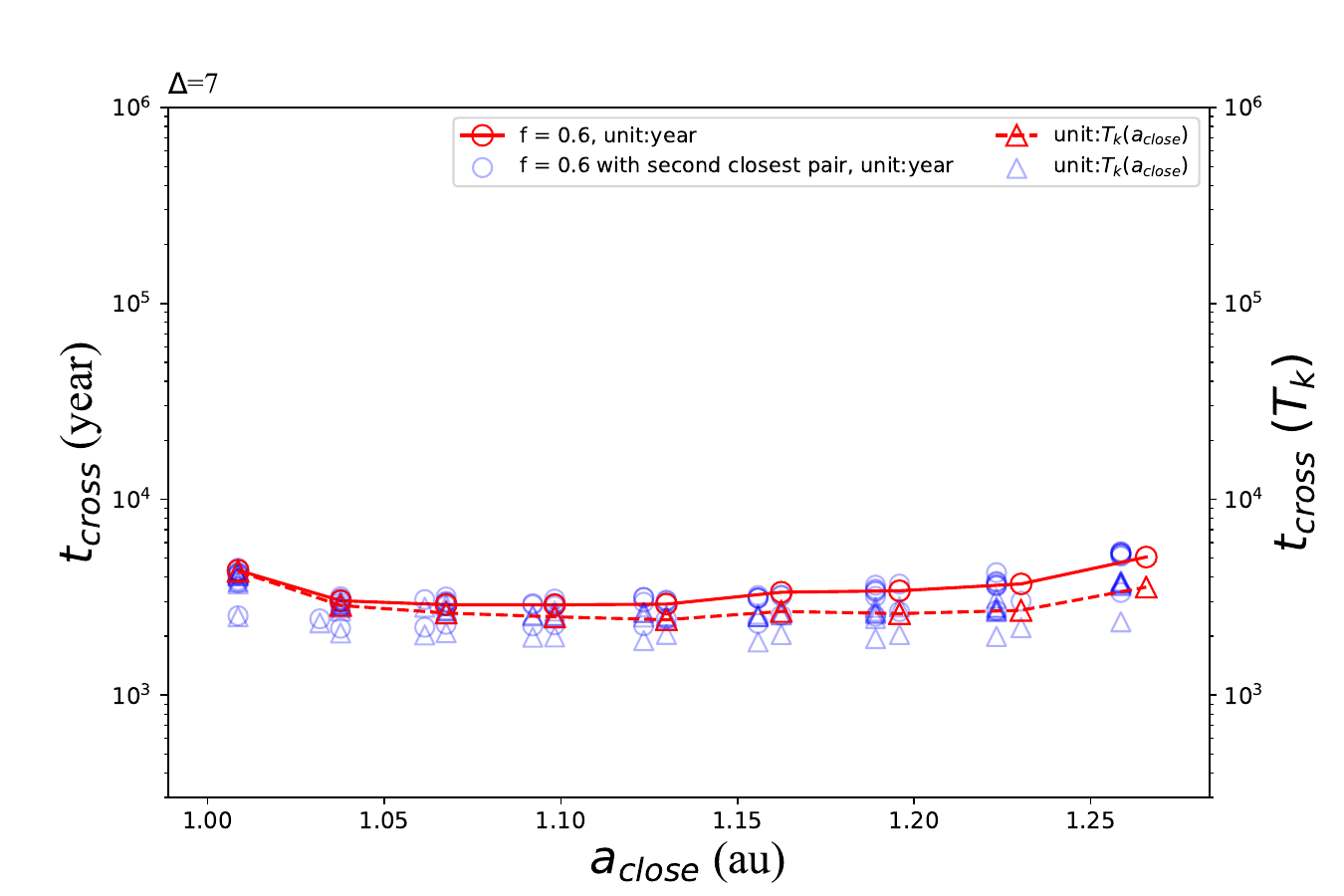}
 \caption{
 Same as Figure\,1(d) for $f=0.6$ in the main text, but for systems with second closest pairs with a second closest pair with a separation of $f=0.8$.
The red symbol represents the same as in Figure\,1(d).
With $a_{\rm close}$ set to a specific value for the closest pair ($f=0.6$), each blue symbol indicates the results of shifting the position of the second closest pair ($f=0.8$).
In the case with the second closest pair, there are cases where the orbit crossing time is slightly shorter than in Figure\,1(d). Meanwhile, the finding in Figure\,1(d) (the orbit crossing time can be expressed by the orbital periods of the closest separation pair) seems to be unchanged.}
\label{fig:pic_supp_2}
\end{figure*}
\printcredits

\bibliographystyle{cas-model2-names}
\bibliography{main}

\begin{thebibliography}{19}
\expandafter\ifx\csname natexlab\endcsname\relax\def\natexlab#1{#1}\fi
\providecommand{\url}[1]{\texttt{#1}}
\providecommand{\href}[2]{#2}
\providecommand{\path}[1]{#1}
\providecommand{\DOIprefix}{doi:}
\providecommand{\ArXivprefix}{arXiv:}
\providecommand{\URLprefix}{URL: }
\providecommand{\Pubmedprefix}{pmid:}
\providecommand{\doi}[1]{\href{http://dx.doi.org/#1}{\path{#1}}}
\providecommand{\Pubmed}[1]{\href{pmid:#1}{\path{#1}}}
\providecommand{\bibinfo}[2]{#2}
\ifx\xfnm\relax \def\xfnm[#1]{\unskip,\space#1}\fi
\bibitem[{Chambers et~al.(1996)Chambers, Wetherill and
  Boss}]{chambers_stability_1996}
\bibinfo{author}{Chambers, J.E.}, \bibinfo{author}{Wetherill, G.W.},
  \bibinfo{author}{Boss, A.P.}, \bibinfo{year}{1996}.
\newblock \bibinfo{title}{The {Stability} of {Multi}-{Planet} {Systems}}.
\newblock \bibinfo{journal}{Icarus} \bibinfo{volume}{119},
  \bibinfo{pages}{261--268}.
\newblock \URLprefix
  \url{https://www.sciencedirect.com/science/article/pii/S0019103596900196},
  \DOIprefix\doi{10.1006/icar.1996.0019}.
\bibitem[{Gratia and Lissauer(2021)}]{GRATIA2021114038}
\bibinfo{author}{Gratia, P.}, \bibinfo{author}{Lissauer, J.J.},
  \bibinfo{year}{2021}.
\newblock \bibinfo{title}{Eccentricities and the stability of closely-spaced
  five-planet systems}.
\newblock \bibinfo{journal}{Icarus} \bibinfo{volume}{358},
  \bibinfo{pages}{114038}.
\newblock \URLprefix
  \url{https://www.sciencedirect.com/science/article/pii/S0019103520303948},
  \DOIprefix\doi{https://doi.org/10.1016/j.icarus.2020.114038}.
\bibitem[{Ida et~al.(2013)Ida, Lin and Nagasawa}]{ida_toward_2013}
\bibinfo{author}{Ida, S.}, \bibinfo{author}{Lin, D.N.C.},
  \bibinfo{author}{Nagasawa, M.}, \bibinfo{year}{2013}.
\newblock \bibinfo{title}{{TOWARD} {A} {DETERMINISTIC} {MODEL} {OF} {PLANETARY}
  {FORMATION}. {VII}. {ECCENTRICITY} {DISTRIBUTION} {OF} {GAS} {GIANTS}}.
\newblock \bibinfo{journal}{ApJ} \bibinfo{volume}{775}, \bibinfo{pages}{42}.
\newblock \URLprefix \url{https://dx.doi.org/10.1088/0004-637X/775/1/42},
  \DOIprefix\doi{10.1088/0004-637X/775/1/42}. \bibinfo{note}{publisher: The
  American Astronomical Society}.
\bibitem[{Izidoro et~al.(2017)Izidoro, Ogihara, Raymond, Morbidelli, Pierens,
  Bitsch, Cossou and Hersant}]{izidoro_breaking_2017}
\bibinfo{author}{Izidoro, A.}, \bibinfo{author}{Ogihara, M.},
  \bibinfo{author}{Raymond, S.N.}, \bibinfo{author}{Morbidelli, A.},
  \bibinfo{author}{Pierens, A.}, \bibinfo{author}{Bitsch, B.},
  \bibinfo{author}{Cossou, C.}, \bibinfo{author}{Hersant, F.},
  \bibinfo{year}{2017}.
\newblock \bibinfo{title}{Breaking the chains: hot super-{Earth} systems from
  migration and disruption of compact resonant chains}.
\newblock \bibinfo{journal}{Monthly Notices of the Royal Astronomical Society}
  \bibinfo{volume}{470}, \bibinfo{pages}{1750--1770}.
\newblock \URLprefix \url{https://doi.org/10.1093/mnras/stx1232},
  \DOIprefix\doi{10.1093/mnras/stx1232}.
\bibitem[{Kokubo and Ida(1998)}]{kokubo_oligarchic_1998}
\bibinfo{author}{Kokubo, E.}, \bibinfo{author}{Ida, S.}, \bibinfo{year}{1998}.
\newblock \bibinfo{title}{Oligarchic {Growth} of {Protoplanets}}.
\newblock \bibinfo{journal}{Icarus} \bibinfo{volume}{131},
  \bibinfo{pages}{171--178}.
\newblock \URLprefix
  \url{https://www.sciencedirect.com/science/article/pii/S0019103597958401},
  \DOIprefix\doi{10.1006/icar.1997.5840}.
\bibitem[{Lissauer et~al.(2011)Lissauer, Ragozzine, Fabrycky, Steffen, Ford,
  Jenkins, Shporer, Holman, Rowe, Quintana, Batalha, Borucki, Bryson, Caldwell,
  Carter, Ciardi, Dunham, Fortney, Gautier, Howell, Koch, Latham, Marcy,
  Morehead and Sasselov}]{lissauer_architecture_2011}
\bibinfo{author}{Lissauer, J.J.}, \bibinfo{author}{Ragozzine, D.},
  \bibinfo{author}{Fabrycky, D.C.}, \bibinfo{author}{Steffen, J.H.},
  \bibinfo{author}{Ford, E.B.}, \bibinfo{author}{Jenkins, J.M.},
  \bibinfo{author}{Shporer, A.}, \bibinfo{author}{Holman, M.J.},
  \bibinfo{author}{Rowe, J.F.}, \bibinfo{author}{Quintana, E.V.},
  \bibinfo{author}{Batalha, N.M.}, \bibinfo{author}{Borucki, W.J.},
  \bibinfo{author}{Bryson, S.T.}, \bibinfo{author}{Caldwell, D.A.},
  \bibinfo{author}{Carter, J.A.}, \bibinfo{author}{Ciardi, D.},
  \bibinfo{author}{Dunham, E.W.}, \bibinfo{author}{Fortney, J.J.},
  \bibinfo{author}{Gautier, T.N.}, \bibinfo{author}{Howell, S.B.},
  \bibinfo{author}{Koch, D.G.}, \bibinfo{author}{Latham, D.W.},
  \bibinfo{author}{Marcy, G.W.}, \bibinfo{author}{Morehead, R.C.},
  \bibinfo{author}{Sasselov, D.}, \bibinfo{year}{2011}.
\newblock \bibinfo{title}{{ARCHITECTURE} {AND} {DYNAMICS} {OF} {KEPLER}'{S}
  {CANDIDATE} {MULTIPLE} {TRANSITING} {PLANET} {SYSTEMS}}.
\newblock \bibinfo{journal}{ApJS} \bibinfo{volume}{197}, \bibinfo{pages}{8}.
\newblock \URLprefix \url{https://dx.doi.org/10.1088/0067-0049/197/1/8},
  \DOIprefix\doi{10.1088/0067-0049/197/1/8}. \bibinfo{note}{publisher: The
  American Astronomical Society}.
\bibitem[{Matsumoto et~al.(2020)Matsumoto, Gu, Kokubo, Oshino and
  Omiya}]{matsumoto_ejection_2020}
\bibinfo{author}{Matsumoto, Y.}, \bibinfo{author}{Gu, P.G.},
  \bibinfo{author}{Kokubo, E.}, \bibinfo{author}{Oshino, S.},
  \bibinfo{author}{Omiya, M.}, \bibinfo{year}{2020}.
\newblock \bibinfo{title}{Ejection of close-in super-{Earths} around low-mass
  stars in the giant impact stage}.
\newblock \bibinfo{journal}{A\&A} \bibinfo{volume}{642}, \bibinfo{pages}{A23}.
\newblock \URLprefix
  \url{https://www.aanda.org/articles/aa/abs/2020/10/aa38332-20/aa38332-20.html},
  \DOIprefix\doi{10.1051/0004-6361/202038332}. \bibinfo{note}{publisher: EDP
  Sciences}.
\bibitem[{Ogihara and Ida(2009)}]{ogihara_n-body_2009}
\bibinfo{author}{Ogihara, M.}, \bibinfo{author}{Ida, S.}, \bibinfo{year}{2009}.
\newblock \bibinfo{title}{N-{BODY} {SIMULATIONS} {OF} {PLANETARY} {ACCRETION}
  {AROUND} {M} {DWARF} {STARS}}.
\newblock \bibinfo{journal}{ApJ} \bibinfo{volume}{699}, \bibinfo{pages}{824}.
\newblock \URLprefix \url{https://dx.doi.org/10.1088/0004-637X/699/1/824},
  \DOIprefix\doi{10.1088/0004-637X/699/1/824}. \bibinfo{note}{publisher: The
  American Astronomical Society}.
\bibitem[{Ogihara et~al.(2018)Ogihara, Kokubo, Suzuki and
  Morbidelli}]{ogihara_formation_2018}
\bibinfo{author}{Ogihara, M.}, \bibinfo{author}{Kokubo, E.},
  \bibinfo{author}{Suzuki, T.K.}, \bibinfo{author}{Morbidelli, A.},
  \bibinfo{year}{2018}.
\newblock \bibinfo{title}{Formation of close-in super-{Earths} in evolving
  protoplanetary disks due to disk winds}.
\newblock \bibinfo{journal}{A\&A} \bibinfo{volume}{615}, \bibinfo{pages}{A63}.
\newblock \URLprefix
  \url{https://www.aanda.org/articles/aa/abs/2018/07/aa32720-18/aa32720-18.html},
  \DOIprefix\doi{10.1051/0004-6361/201832720}. \bibinfo{note}{publisher: EDP
  Sciences}.
\bibitem[{Petit et~al.(2020)Petit, Pichierri, Davies and
  Johansen}]{petit_path_2020}
\bibinfo{author}{Petit, A.C.}, \bibinfo{author}{Pichierri, G.},
  \bibinfo{author}{Davies, M.B.}, \bibinfo{author}{Johansen, A.},
  \bibinfo{year}{2020}.
\newblock \bibinfo{title}{The path to instability in compact multi-planetary
  systems}.
\newblock \bibinfo{journal}{A\&A} \bibinfo{volume}{641}, \bibinfo{pages}{A176}.
\newblock \URLprefix
  \url{https://www.aanda.org/articles/aa/abs/2020/09/aa38764-20/aa38764-20.html},
  \DOIprefix\doi{10.1051/0004-6361/202038764}. \bibinfo{note}{publisher: EDP
  Sciences}.
\bibitem[{Pu and Wu(2015)}]{pu_spacing_2015}
\bibinfo{author}{Pu, B.}, \bibinfo{author}{Wu, Y.}, \bibinfo{year}{2015}.
\newblock \bibinfo{title}{{SPACING} {OF} {KEPLER} {PLANETS}: {SCULPTING} {BY}
  {DYNAMICAL} {INSTABILITY}}.
\newblock \bibinfo{journal}{ApJ} \bibinfo{volume}{807}, \bibinfo{pages}{44}.
\newblock \URLprefix \url{https://dx.doi.org/10.1088/0004-637X/807/1/44},
  \DOIprefix\doi{10.1088/0004-637X/807/1/44}. \bibinfo{note}{publisher: The
  American Astronomical Society}.
\bibitem[{{Quillen}(2011)}]{2011MNRAS.418.1043Q}
\bibinfo{author}{{Quillen}, A.C.}, \bibinfo{year}{2011}.
\newblock \bibinfo{title}{{Three-body resonance overlap in closely spaced
  multiple-planet systems}}.
\newblock \bibinfo{journal}{mnras} \bibinfo{volume}{418},
  \bibinfo{pages}{1043--1054}.
\newblock \DOIprefix\doi{10.1111/j.1365-2966.2011.19555.x},
  \href{http://arxiv.org/abs/1106.0156}{\tt arXiv:1106.0156}.
\bibitem[{Rein and Liu(2012)}]{rein_rebound_2012}
\bibinfo{author}{Rein, H.}, \bibinfo{author}{Liu, S.F.}, \bibinfo{year}{2012}.
\newblock \bibinfo{title}{{REBOUND}: an open-source multi-purpose {N}-body code
  for collisional dynamics}.
\newblock \bibinfo{journal}{A\&A} \bibinfo{volume}{537}, \bibinfo{pages}{A128}.
\newblock \URLprefix
  \url{https://www.aanda.org/articles/aa/abs/2012/01/aa18085-11/aa18085-11.html},
  \DOIprefix\doi{10.1051/0004-6361/201118085}. \bibinfo{note}{publisher: EDP
  Sciences}.
\bibitem[{Rein and Spiegel(2015)}]{rein_ias15_2015}
\bibinfo{author}{Rein, H.}, \bibinfo{author}{Spiegel, D.S.},
  \bibinfo{year}{2015}.
\newblock \bibinfo{title}{ias15: a fast, adaptive, high-order integrator for
  gravitational dynamics, accurate to machine precision over a billion orbits}.
\newblock \bibinfo{journal}{Monthly Notices of the Royal Astronomical Society}
  \bibinfo{volume}{446}, \bibinfo{pages}{1424--1437}.
\newblock \URLprefix \url{https://doi.org/10.1093/mnras/stu2164},
  \DOIprefix\doi{10.1093/mnras/stu2164}.
\bibitem[{Rein and Tamayo(2015)}]{rein_whfast_2015}
\bibinfo{author}{Rein, H.}, \bibinfo{author}{Tamayo, D.}, \bibinfo{year}{2015}.
\newblock \bibinfo{title}{whfast: a fast and unbiased implementation of a
  symplectic {Wisdom}–{Holman} integrator for long-term gravitational
  simulations}.
\newblock \bibinfo{journal}{Monthly Notices of the Royal Astronomical Society}
  \bibinfo{volume}{452}, \bibinfo{pages}{376--388}.
\newblock \URLprefix \url{https://doi.org/10.1093/mnras/stv1257},
  \DOIprefix\doi{10.1093/mnras/stv1257}.
\bibitem[{Smith and Lissauer(2009)}]{smith_orbital_2009}
\bibinfo{author}{Smith, A.W.}, \bibinfo{author}{Lissauer, J.J.},
  \bibinfo{year}{2009}.
\newblock \bibinfo{title}{Orbital stability of systems of closely-spaced
  planets}.
\newblock \bibinfo{journal}{Icarus} \bibinfo{volume}{201},
  \bibinfo{pages}{381--394}.
\newblock \URLprefix
  \url{https://www.sciencedirect.com/science/article/pii/S0019103508004570},
  \DOIprefix\doi{10.1016/j.icarus.2008.12.027}.
\bibitem[{Wu et~al.(2019)}]{wu_dynamical_2019}
\bibinfo{author}{Wu}, et~al., \bibinfo{year}{2019}.
\newblock \bibinfo{title}{Dynamical instability and its implications for
  planetary system architecture}.
\newblock \bibinfo{journal}{Monthly Notices of the Royal Astronomical Society}
  \bibinfo{volume}{484}, \bibinfo{pages}{1538--1548}.
\newblock \URLprefix \url{https://doi.org/10.1093/mnras/stz054},
  \DOIprefix\doi{10.1093/mnras/stz054}.
\bibitem[{Yoshinaga et~al.(1999)Yoshinaga, Kokubo and
  Makino}]{yoshinaga_stability_1999}
\bibinfo{author}{Yoshinaga, K.}, \bibinfo{author}{Kokubo, E.},
  \bibinfo{author}{Makino, J.}, \bibinfo{year}{1999}.
\newblock \bibinfo{title}{The {Stability} of {Protoplanet} {Systems}}.
\newblock \bibinfo{journal}{Icarus} \bibinfo{volume}{139},
  \bibinfo{pages}{328--335}.
\newblock \URLprefix
  \url{https://www.sciencedirect.com/science/article/pii/S0019103599960980},
  \DOIprefix\doi{10.1006/icar.1999.6098}.
\bibitem[{Zhou et~al.(2007)Zhou, Lin and Sun}]{zhou_post-oligarchic_2007}
\bibinfo{author}{Zhou, J.L.}, \bibinfo{author}{Lin, D.N.C.},
  \bibinfo{author}{Sun, Y.S.}, \bibinfo{year}{2007}.
\newblock \bibinfo{title}{Post-oligarchic {Evolution} of {Protoplanetary}
  {Embryos} and the {Stability} of {Planetary} {Systems}}.
\newblock \bibinfo{journal}{ApJ} \bibinfo{volume}{666}, \bibinfo{pages}{423}.
\newblock \URLprefix
  \url{https://iopscience.iop.org/article/10.1086/519918/meta},
  \DOIprefix\doi{10.1086/519918}. \bibinfo{note}{publisher: IOP Publishing}.

\end{thebibliography}

\end{document}